\documentclass[fleqn,usenatbib]{mnras}
\usepackage{Preamble}
\usepackage{xcolor}
\usepackage[T1]{fontenc}
\usepackage{hyperref}
\usepackage{longtable}

\DeclareUnicodeCharacter{2212}{-}

\newcommand{\uat}[1]{#1}
\usepackage{newtxtext,newtxmath}
\usepackage[T1]{fontenc}
\DeclareRobustCommand{\VAN}[3]{#2}
\let\VANthebibliography\thebibliography
\def\thebibliography{\DeclareRobustCommand{\VAN}[3]{##3}\VANthebibliography}
\usepackage{graphicx}	% Including figure files
\usepackage{amsmath}	% Advanced maths commands
\usepackage{multirow}
\usepackage{multicol}
\usepackage{ulem}
\title[QPO in blazar PKS 0735+178]{Detection of  Quasiperiodic Oscillations in the Blazar PKS 0735+178 with TESS}  
\author[Kishore et al.]{Shubham Kishore$^{1}$ \thanks{E-mail: amp700151@gmail.com (SK)}\orcidlink{0000-0001-8716-9412},
Alok C. Gupta$^{2}$\orcidlink{0000-0002-9331-4388},
Paul J. Wiita$^{3}$\orcidlink{0000-0002-1029-3746},
Sandeep K. Mondal$^{4}$\thanks{E-mail: sandeep@sjtu.edu.cn}\orcidlink{0000-0003-2445-9935}
M. Vivek$^{1}$\orcidlink{0000-0001-5937-331X}
\\
\\
$^{1}$Indian Institute of Astrophysics, 2nd Block, Koramangala, Bangalore, 560034, India\\
$^{2}$Aryabhatta Research Institute of Observational Sciences (ARIES), Manora Peak, Nainital 263001, India\\
$^{3}$Department of Physics, The College of New Jersey, 2000 Pennington Rd., Ewing, NJ 08628-0718, USA\\
$^{4}$Tsung-Dao Lee Institute, Shanghai Jiao Tong University, 1 Lisuo Road, Shanghai, 201210, People’s Republic of China}
\date{Accepted  XXXX. Received 2026 April 30; in original form 2026 April 27}
\pubyear{2026}
\begin{document}
\label{firstpage}
\pagerange{\pageref{firstpage}--\pageref{lastpage}}
\maketitle
\begin{abstract}
We report here the detection of signatures of a quasiperiodic oscillation (QPO) and a short flare in the optical light curve of the blazar PKS~0735+178, observed in two sectors, 71 and 72, spanning around 49 days with the Transiting Exoplanet Survey Satellite. The modest flare in sector 71 lasted $\sim4.3$~days and appears as a combination of two sub-flares.
In sector~72, a transient QPO with a period $\sim11.2$~hours is detected at local and global significance levels of $4.11\sigma~\text{and}~3.06\sigma$, respectively. We used weighted wavelet z transform, Lomb-Scargle periodogram, and phase dispersion minimization analysis techniques to look for and confirm the QPO feature.
We also performed a segment-wise statistical inspection of these light curves and discuss here possible mechanisms that could explain the observed flux behavior.
\end{abstract}
\section{Introduction}\label{sec:intro}
\noindent
A subclass of radio-loud (RL) active galactic nuclei (AGN), blazars emit radiation in all electromagnetic (EM) bands, from radio to $\gamma-$rays. Blazars contain relativistic charged particle jets that are nearly aligned with the observer's line of sight \citep{1995PASP..107..803U}. The jet emission dominates the total observed fluxes from blazars in almost all EM bands due to relativistic effects, which produce significantly magnified observed emissions due to the extremely small viewing angle. Blazars exhibit notable flux, spectral, and polarization variation across the whole EM spectrum, with non-thermal processes dominating their emission \citep[e.g.][and references therein]{1997ApJ...486..799U, 2007ApJ...670..968B, 2009ApJ...696L.150A, 2010Natur.463..919A, 2010ApJ...721.1425A, 2015ApJ...807...79H, 2017MNRAS.472..788G, 2023ApJ...957L..11G, 2017MNRAS.466.3762R, 2024A&A...692A..48R}. Flat spectrum radio quasars (FSRQs) and BL Lacertae objects (BLLs) are together referred to as blazars. In the composite optical/UV spectrum, BLLs show featureless or very faint emission lines with equivalent widths (EWs) $\leq5$~\AA, whereas FSRQs show strong emission lines. Blazar spectral energy distributions (SEDs) exhibit a double-hump structure, with the low-energy and high-energy humps peaking in infrared (IR) to X-ray bands and $\gamma-$ray energies, respectively \citep[e.g.][]{1998MNRAS.299..433F}. Ultra-relativistic electrons in the jet are the source of the low-energy component of the SED via synchrotron radiation, and the high-energy component can usually be explained by inverse Compton (IC) emission, either via synchrotron self-Compton or external Compton processes \citep[e.g.][and references therein]{2013ApJ...768...54B,2019ARA&A..57..467B,2025A&ARv..33....8R}. \\
\\
The light curves (LCs) of X-ray binaries have frequently shown nearly periodic oscillations, also known as quasi-periodic oscillations (QPOs) \citep{2006ARA&A..44...49R}. The literature has reported the detection of QPOs in blazars and other subclasses of AGN with diverse periods, despite the fact that the LCs of AGN are primarily non-periodic across the whole EM spectrum \citep[e.g.][and references therein]{1985Natur.314..146C, 1985Natur.314..148V, 1989ApJ...347..171F, 1993Natur.361..233P, 1998MNRAS.295L..20I}. After evidence of a QPO with a period of $\sim$1 hour was discovered in the X-ray LC of the AGN RE J1034+396 \citep{2008Natur.455..369G}, the search for QPOs in various subclasses of AGN LCs has emerged as an important scientific project in extragalactic astronomy that has employed various analysis techniques. There have been strong claims of occasional QPO detection in several blazars on diverse timescales ranging from a few minutes to days to weeks to months and even years in different EM bands, e.g., radio \citep[][and references therein]{2013MNRAS.436L.114K, 2014MNRAS.443...58W, 2024ApJ...977..166T, 2026A&A...707A.371G}, infrared (IR)/optical \citep[][and references therein]{2009ApJ...690..216G, 2014ApJ...793L...1S, 2016AJ....151...54S, 2016ApJ...832...47B, 2022ApJ...934....3Z}, X-ray \citep[][and references therein]{2009A&A...506L..17L, 2009ApJ...696.2170R, 2023ApJ...950..174S, 2025MNRAS.541.3008Z}, and $\gamma-$rays \citep[][and references therein]{2018NatCo...9.4599Z, 2019MNRAS.487.3990B, 2022MNRAS.510.3641R, 2023ApJ...950..173D, 2025MNRAS.541.2955P}. There even have been some claims of blazar QPOs being detected simultaneously in more than one EM band \citep[e.g.][and references therein]{2015ApJ...813L..41A, 2020A&A...642A.129S, 2021MNRAS.501...50S, 2022Natur.609..265J}. Evidence for the detection of QPOs in a few other, less variable, subclasses of AGN has also been reported \citep[e.g.][and references therein]{2008Natur.455..369G, 2014MNRAS.445L..16A, 2015MNRAS.449..467A, 2016ApJ...819L..19P, 2018A&A...616L...6G, 2025ApJ...992L..13Z}. \\
\\
Detection of flares in blazars on diverse timescales in different EM bands is very common. In general, blazars can be said to be in one of three different flux states, namely, quiescent, pre- and/or post-flare, and outbursting or flaring \citep[e.g.][and references therein]{1996A&A...305L..17S, 2013MNRAS.436.1530R, 2022Natur.609..265J, 2025ApJS..276....1D, MANZOOR2026100619}. In flare and outburst states, the emission is due to nonthermal processes, but in the quiescent states, the observed flux of a blazar can be a mixture of thermal and nonthermal emission. There have been quite a few detections of flares in many blazars in different EM bands on diverse timescales \citep[e.g.][and references therein]{2006A&A...451..435M, 2007ApJ...664L..71A, 2010ApJ...718..279G, 2017ApJ...841..123P, 2022Natur.609..265J, 2024A&A...690A.223K, 2024ApJ...960...11K}. In simultaneous multi-wavelength (MW) observations of blazars flares are typically detected in different EM bands with different amplitudes, with interband temporal lags varying from essentially zero to a few hundreds of days \citep[e.g.][and references therein]{1997ApJ...486..799U, 2006A&A...456..911G, 2007A&A...473..819R, 2009ApJ...696L.150A, 2010ApJ...715..362J, 2013MNRAS.436.1530R, 2024A&A...692A..48R}. \\
\\
PKS 0735+178 is a BLL at redshift $z = 0.45\pm 0.06$, as estimated from the detection of its host galaxy \citep{2012A&A...547A...1N}. Using about a century (1906--2001) of optical data, \citet{2004PASP..116..161Q} reported a dominant QPO period of 13.8 years. The source has been extensively studied in search of its optical flux, color, and polarization variability and has shown variations of these physical properties on diverse timescales \citep[e.g.][and references therein]{2007A&A...467..465C, 2009MNRAS.399.1622G, 2010A&A...517A..63T, 2021PASP..133g4101Y, 2024MNRAS.528.4702M}. It has also shown significant flux and spectral variations in its MW observations \citep[e.g.][and references therein]{2017ApJ...837..127G, 2022ApJ...933..224F, 2023ApJ...954...70A, 2024MNRAS.527.8746P, 2024MNRAS.529.3503B}. Several neutrino events detected by the IceCube, Baikal, Baksan, and KM3NeT neutrino telescopes during the blazar PKS 0735+178's December 2021 flare in the $\gamma-$ray, X-ray, ultraviolet, and optical bands may be linked to this object. This blazar was thoroughly studied in order to look into the spectral and temporal variations in the MW emission during the neutrino events \citep{2023MNRAS.519.1396S, 2023ApJ...954...70A, 2024MNRAS.529.3503B}. Superluminal motion has been detected by very long baseline interferometry (VLBI) studies of PKS 0735+178’s parsec-scale jet, yielding estimates of the jet viewing angle, $\Theta = \rm{5.7}\pm\rm{1.4}$ degrees, and Doppler factor $\delta = \rm{8.5}\pm\rm{4.0}$ \citep{2017ApJ...846...98J}. \\
\\
In this paper, we report the detection of a two-component flare spanning around 4.5 days in composite, and a QPO with a period of around 11.2 hours in the Transiting Exoplanet Survey Satellite (TESS) optical LC of the blazar PKS 0735+178. Variability with timescales of less than a day is often called intraday variability (IDV) \citep{1995ARA&A..33..163W} and is one of the most puzzling issues in blazar astrophysics. \\ 
\\
In Section 2, we provide information about the TESS optical data and Fermi $\gamma-$ray data we have investigated. In Section 3, different analysis techniques that we implemented on the optical data are presented. In Section 4, we present the results with the optical data. The $\gamma$-ray data show an almost invariable behavior, so we give only a qualitative comparison between these bands in Section 4. A discussion and conclusions are presented in Section 5. 
\begin{figure}
    \includegraphics[width=0.95\linewidth, trim=.43cm .9cm 0 0, clip ]{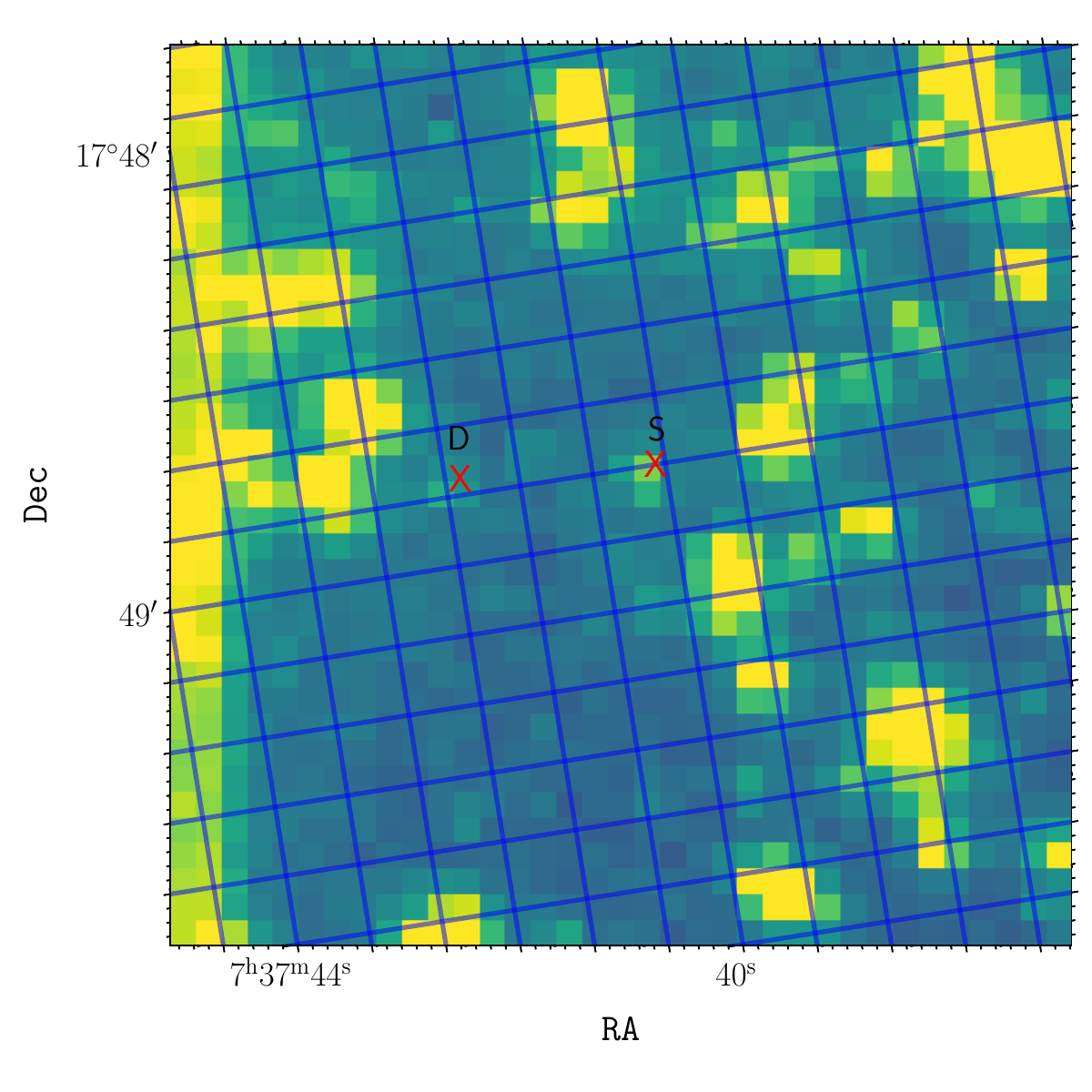}
    \caption{A $35\times35$ pixel cutout of sector 71 TESS field of view, including the source (`S') and the comparison star (`D')}
    \label{fig:1}
\end{figure}
\begin{figure*}
    \centering
    \includegraphics[width=0.9\linewidth]{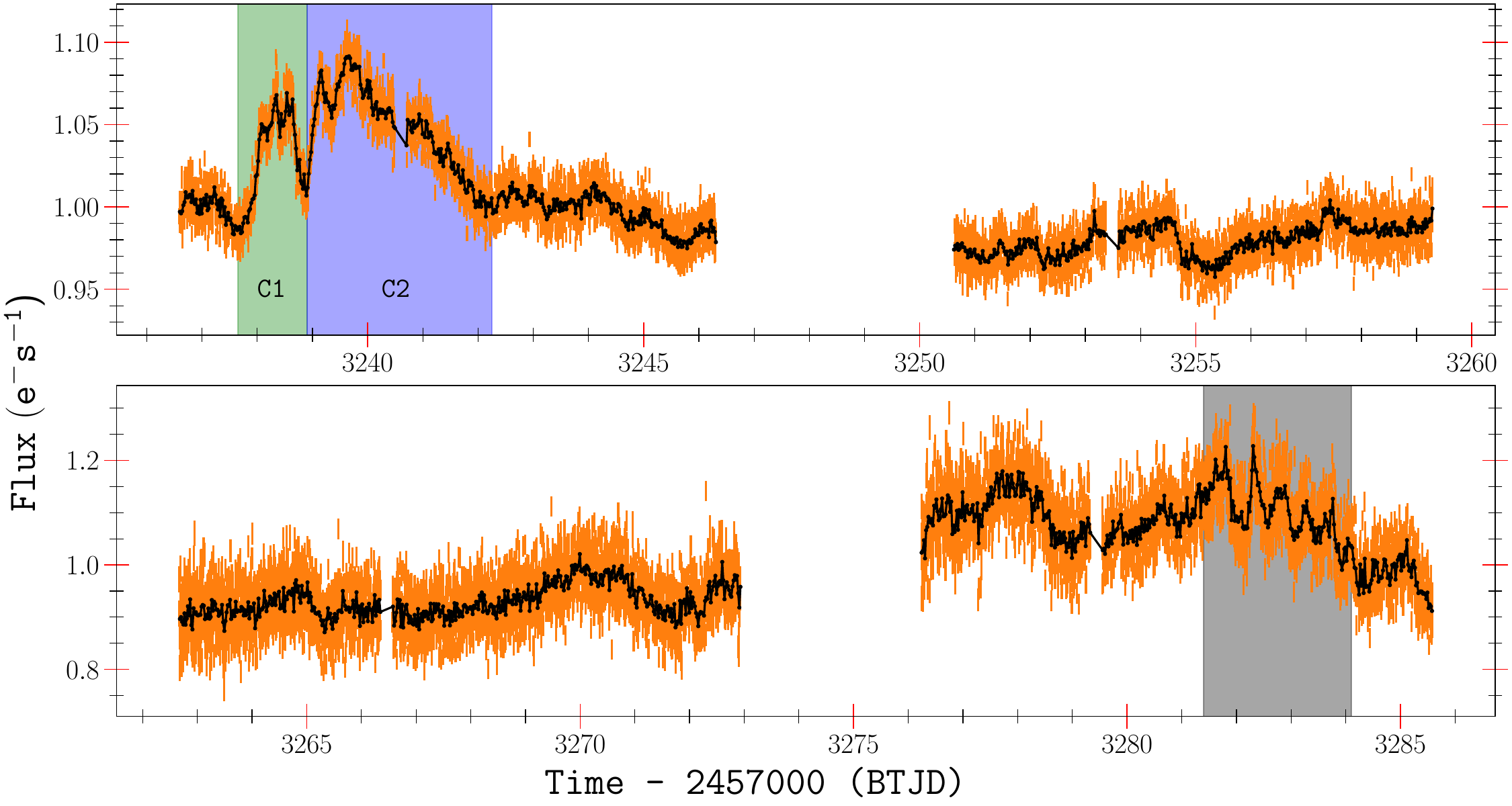}
    \caption{Normalized LCs of PKS 0735+178 observed in sectors 71 (upper panel) and 72 (lower panel); orange colored points have a cadence of 200s which we have used in our analysis, black points are 0.5~hours binned LCs, overplotted here to demonstrate the flux variations more clearly. Shaded regions C1 and C2 denote the two components of the strongest flare during these observations; the shaded region in sector 72 LC highlights the span where the QPO search  was conducted.}
    \label{fig:2}
\end{figure*}
 \begin{figure*}
     \centering
     \includegraphics[width=0.9\linewidth]{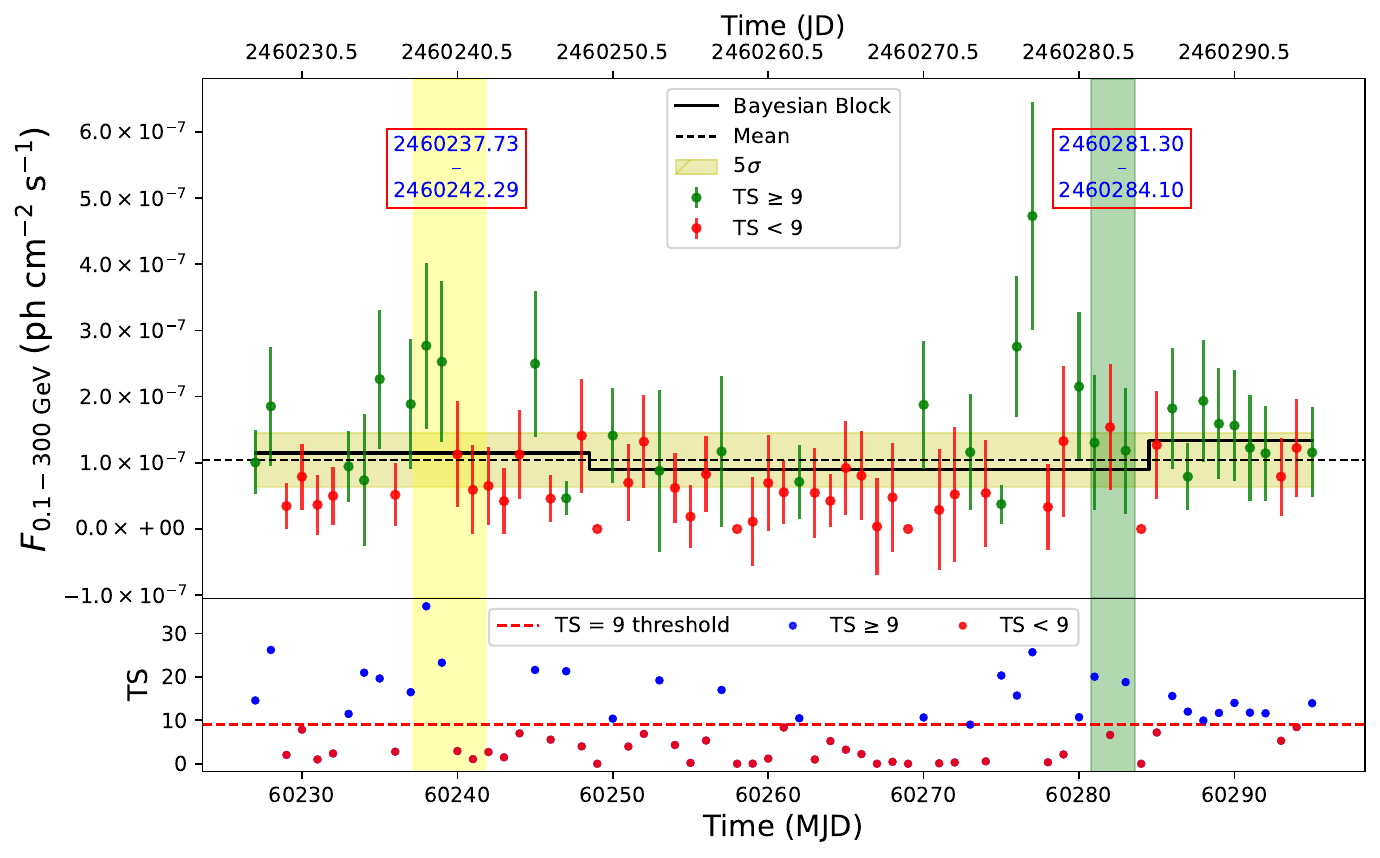}
     \caption{$\gamma-$ray LC of PKS~0735+178 during the time window of TESS's sectors 71 and 72; the yellow and green shaded regions respectively mark the epochs of the optical flare and QPO. Data points are color-coded by statistical significance: red indicates TS$<$9, while significant detections (TS$\geq$9) are shown in green (upper panel) and blue (lower panel). The horizontal red-dashed line in the lower panel marks the TS=9 threshold.}
     \label{fig:3}
 \end{figure*}
\section{Data acquisition}
\subsection{Optical dataset}
\noindent
TESS observes in a survey mode, covering the whole sky through a total of 26 sectors. It continues to observe each sector for $\sim27$ days \citep{2016SPIE.9913E..3EJ}. We obtained an $11\times11$ pixel cutout of the full frame images (FFI) of the portion of sky pointing towards PKS~0735+178, centered at the source, or `S', which was observed in sectors 71 and 72 (19\textsuperscript{th} October 2023 to 7\textsuperscript{th} December 2023). Additionally, we obtained a similar cutout of the FFI of the comparison star `D' during sectors 71 and 72 to perform and obtain normalized differential photometric LCs of PKS~0735+178 \citep[see Fig.~1 of][for the field of PKS~0735+178 and the nearby comparison star `D']{1985AJ.....90.1184S}. The star `D' was selected for the differential photometry because of its being relatively less crowded than other possible comparison stars, so we could neglect other flux contributions from its vicinity.\\
\\
For differential photometry, an optimal aperture was chosen for both  PKS~0735+178 and star `D' at the center of the respective cutouts, such that the pixel flux value exceeds the background by at least $2\sigma$ levels, after which aperture photometric LCs were obtained for the source and `D' in each sector. Since the source and `D' are not very distant ($(\Delta\alpha,~\Delta\delta)=(0.003862^\circ, 0.044377^\circ)$; Fig.~\ref{fig:1} demonstrates the positions of the source and `D' in a $35\times35$ cutout of the full frame image in sector 71), it is safe to assume that any instrumental artifact would affect PKS~0735+178 and `D' in a similar manner. We then obtained the background LCs to obtain background-subtracted LCs of the source and `D' in the two sectors. Each of the sectoral `D' LCs was then divided by its corresponding median flux value to make a normalized `D' LC. Since the comparison star remains invariable, the behavior of the normalized `D' LC mimics the pixel response function throughout the sectoral observation, and should ideally be unity. However, near the ends of each segment of both sectors, this normalized LC shows relatively high values, so we further made a cutoff of $15\%$ about unity to the pixel response function values, along with additional cuts where they show a sharp change, to discard any contingent behavior of the pixels, and eliminate those data points from the `D' and `S' LCs. \\
\\
The background-subtracted LCs of the source are then divided by these pixel response function values at corresponding epochs, followed by a further division by the mean sectoral flux counts of the source to obtain its sectoral normalized differential photometric LCs. In the appendix, we present the background LC, the normalized LC of `D', the raw LC of PKS~0735+178, and the pixel response function corrected LC of the source in sectors 71 and 72 (Fig.~\ref{fig:Ap1} and \ref{fig:Ap2}). The sectoral normalized LCs of the source in Fig.~\ref{fig:2} are simply obtained by dividing the pixel response function corrected LC by their sectoral mean values. The data acquisition and handling were performed using the {\tt lightkurve} \citep{2018ascl.soft12013L} Python package. Fig.~\ref{fig:2} presents the thus obtained normalized LCs of the source PKS~0735+178 in the two sectors 71 and 72. The TESS observations have a default gap of $2-4$~days in the middle of each sector during which it sends the observed data to Earth or waits for any instructions to be uploaded, creating a default unevenness in cadences that makes it difficult to search for possible periodicities when an entire sector is considered. Thus, we divided the two sectoral LCs into four segments: 1/71, 2/71, 1/72, and 2/72, for the purpose of our analysis.
\subsection{$\gamma-$ray dataset}
\noindent
For this work, we analyzed \textit{Fermi}-LAT Pass 8 $\gamma$-ray data of PKS~0735+178 obtained from the Fermi Science Support Center (FSSC) server \citep{fermi_lat_query}, covering a period of 69 days (9\textsuperscript{th} October 2023 to 17\textsuperscript{th} December 2023). The data were extracted from a circular region of a radius 30$^\circ$, centered on PKS 0735+178, covering the energy range 0.1–300 GeV. The \textit{Fermi}-LAT is a space-based high-energy gamma-ray instrument capable of observing photons over an energy range between 20 MeV -- 300 GeV. It is an imaging, pair-conversion telescope with a wide-field-of-view (2.7 sr at 1 GeV and above \citep{Atwood_2009}), monitoring the entire sky approximately every three hours in this $\gamma$-ray band. \\
\\
The dataset was analyzed using an open-source Python package Fermipy \citep[v1.0.1;][]{Fermipy_Version}. For analysis, we defined a square region of 10$^\circ \times$10$^\circ$ centered on the source of interest. An energy cut between 0.1 GeV and 300 GeV was applied, while to reduce contamination from the Earth’s limb gamma-rays, a zenith angle cut above 90$^\circ$ was also imposed. For event selection, we used evclass=128 and evtype=3 corresponding to the `P8R3\_SOURCE' class, including both front \& back events. To ensure high-quality data collection, we applied
$ \texttt{DATA\_QUAL} > 0 \ \&\& \ \texttt{LAT\_CONFIG} == 1 $. \\
\\
We used the \textit{Fermi}-LAT Fourth Source Catalog Data Release 4 \citep[4FGL-DR4; gll\_psc\_v35.fits;][]{2023arXiv230712546B}, along with the Galactic diffuse emission described by the gll\_iem\_v07 template \citep{2016ApJS..223...26A} and the isotropic extragalactic background modeled using iso\_p8r3\_source\_v3\_v1.txt. During the likelihood analysis, the parameters of the target source, 4FGL J0738.1+1742, together with the normalizations of the isotropic (isodiff) and Galactic diffuse (galdiff) components, were left free.\\
\\
The target source was modeled with a log-parabola spectrum, with its normalization, spectral index (alpha), and curvature (beta) allowed to vary. In addition, normalizations of all sources within the 5$^\circ$ around 4FGL J0738.1+1742 were kept free. The fit was performed using \texttt{gta.fit()} with the NEWMINUIT optimizer, iterating until the fit quality reached 3 to obtain the best-fit model. After this step, following the Fermipy user documentation \citep{fermipy_docs}, the 1-Day binned \textit{Fermi}-LAT gamma-ray light curve for PKS 0735+178 was extracted as shown in the upper panel of Fig.~\ref{fig:3}. The lower panel shows the Test Statistic (TS). Here, TS stands for Test Statistic. It is the primary numerical value used to quantify the statistical significance of a source detection. TS = 25 corresponds to 5$\sigma$ detection, whereas TS = 9 corresponds to 3$\sigma$ detection. In both panels, red points correspond to TS$<9$, while points with TS$\geq 9$ are shown in green (upper panel) and blue (lower panel). In the upper panel, the horizontal dashed black line represents the average gamma-ray flux, the solid black line shows the Bayesian Block representation \citep{Bayesian_Block_2013}, and the horizontal yellow band indicates the $5\sigma$ range about the mean flux. In the lower panel, the horizontal red-dashed line marks TS = 9. The vertical yellow and green bands denote the epochs of the optical flare and QPO, respectively.
\section{Data Analysis}
\subsection{Fractional Variability}
\noindent
Blazars show diverse variability properties all over the EM spectrum on all timescales ranging from hours to years. When observed over appreciable timespans, they usually show both variable and quiescent phases. The four segmented LCs presented in Fig.~\ref{fig:2} show both quite variable and nearly non-variable states over spans of around 10 days. The root mean square variability amplitude or the square root of excess variance \citep{2003MNRAS.345.1271V} provides a simple way to quantify the intrinsic variation of LCs for any timespan. Taking into account the associated uncertainty, it has the ability to discard the variations due to any instrumental Poisson noise. For any timeseries data, $y(t)$, \\
\begin{equation}
    F_{var}=\sqrt{\frac{S^2-\overline{\sigma_{err}^2}}{\bar{y}^2}},
\end{equation}
\begin{equation}
    S^2=\frac{1}{n-1}\sum(y-\bar{y})^2
\end{equation}
and
\begin{equation}
 (F_{var})_{err}=\sqrt{\left[\sqrt{\frac{1}{2n}}\frac{\overline{\sigma_{err}^2}}{F_{var}\bar{y}^2}\right]^2+\left[\sqrt{\frac{\overline{\sigma_{err}^2}}{n}}\frac{1}{\bar{y}}\right]^2},
\end{equation}\\
here \(F_{var}\) is the fractional root mean square variability amplitude; $S^2-\overline{\sigma_{err}^2}$ is residual variability after removing the expected contribution from the associated uncertainties; \(\bar{y},\ \overline{\sigma_{err}^2}\ \text{and}\ n\) respectively denote the mean flux, mean square uncertainty and the number of data points. We examined the segmented LCs with the above variability test and present them in Table~\ref{tab:2}. Additionally, we examined separately the flaring region of the first segment of sector~71, evaluating $F_{var}$ with respect to a nominal mean flux before the onset of the flare.
  \begin{table*}
     \centering
    \caption{Segmentwise statistical characteristics of the LCs of sectors 71 and 72}
     \begin{tabular}{c c c c c c}\hline
     Segment & Epochs& $F_{var}$ & \multicolumn{3}{c}{Power law parameters}\\
     &(BTJD)& (\%)& $A$ & $\alpha$ & c \\
     \hline
         1/71 & $3236.588~-~3246.304$ & $3.01\pm0.01$ & $1.38_{0.15}^{0.16}\times10^1$ & $2.27_{0.07}^{0.07}$ & $7.30_{0.36}^{0.36}\times10^{-2}$\\
         2/71 & $3250.628~-~3259.289$ & $1.15\pm0.01$ & $1.28_{0.20}^{0.24}\times10^1$ & $2.13_{0.14}^{0.16}$ & $4.88_{0.23}^{0.23}\times10^{-1}$\\
         1/72 & $3262.667~-~3272.930$ & $4.73\pm0.03$ & $6.77_{1.19}^{1.45}$ & $2.01_{0.19}^{0.22}$ & $6.54_{0.28}^{0.28}\times10^{-1}$\\
         2/72 & $3276.238~-~3285.587$ & $6.46\pm0.03$ & $1.89_{0.23}^{0.26}\times10^{1}$ & $1.97_{0.08}^{0.09}$ & $3.38_{0.16}^{0.17}\times10^{-1}$\\
         \hline
     \end{tabular}
     \label{tab:2}
 \end{table*}
 \subsection{Weighted Wavelet Z analysis (WWZ)}
 \noindent
 To inspect for the existence of a periodic or quasi-periodic component in any timeseries data, multiple techniques have been formulated. One commonly employed technique is the WWZ method, which has the ability to dissociate the timeseries into time and frequency domains simultaneously. This method utilizes the convolution of a mother wavelet with the timeseries to produce a two-dimensional map of normalized power as a function of time and temporal frequency, utilizing scale and location parameters of the mother wavelet. The WWZ technique has an advantage over other approaches because it can clearly reveal the length of time any periodic fluctuation persists in the timeseries and is very helpful in detecting and assessing any transient QPOs, which are the type typically seen in blazars. The two-dimensional map can also be utilized to visually segment a specific portion of the timeseries where the periodic fluctuation is maximum and to evaluate its significance. \\
 \\
 We considered the Morlet wavelet, given by $f(z)=e^{-cz^2}(e^{iz}-e^{-1/4c})$,  as the mother wavelet to convolve with the LC, using $z=\omega(t-\tau)$. We then can evaluate the WWZ map, which is given as 
 \begin{equation}
         W[\omega, \tau : x(t_\alpha)] = \omega^{1/2} \sum_{\alpha=1}^{N}x(t_\alpha)f^*[\omega(t_\alpha-\tau)],\\
    \label{DWT_form}
 \end{equation}
 where $x(t_\alpha),~\omega,~\tau,~\text{and}~N$ are the flux at the timestamp $t_\alpha$, angular frequency, wavelet location, and number of data points, respectively. Because of the finite data length, an edge effect, known as the region of influence (ROI) comes into play while evaluating the WWZ map. We employed this test over the oscillatory fluctuations observed in sector~72. As this feature in the second segment of sector 72 appears from BTJD~$\sim$ 3281 to 3284, to encounter the edge effect, we conservatively employed the WWZ analysis between the epochs BTJD~3280.47 and 3285.65. The resultant  map is presented in the upper panel of Fig.~\ref{fig:4}, where the mildly white shaded region denotes the ROI, where the signal is unreliable. 
 \subsection{Lomb-Scargle Periodogram}
 \noindent
 To further confirm the apparent QPO signal against the background power spectrum, we utilized the well-known and traditionally utilized generalized Lomb-Scargle periodogram \citep[e.g.][and references therein]{1976Ap&SS..39..447L, 1982ApJ...263..835S, 2009A&A...496..577Z}. This method allows the estimation of the intrinsic power spectral density (PSD) or periodogram shape of any timeseries data, along with the detection of any periodic behavior in it. A possible QPO appears as a peak in the periodogram. The PSDs of blazars are most commonly described at temporal frequencies ($\nu$) by a red noise power law shape \citep[e.g.][]{2005A&A...431..391V} given as 
 \begin{equation}
 P(\nu)=A\nu^{-\alpha}+c ,
 \label{power_law}
 \end{equation}
 where $A,~\alpha~\text{and}~c$ respectively denote the normalization constant, PSD slope, and white noise level. Other shapes have also been used to describe the PSD shapes on timescales of a few days \citep[e.g.][and references therein]{2026ApJ...998..317K}. In addition to indicating whether any QPOs are present, the PSD is also helpful in extracting information on the source of emission as different physical models for AGN variability can yield distinct ranges for the PSD slopes \citep[e.g.][]{2015ApJ...805...91M, 2016ApJ...820...12P, 2018ApJ...869...32L, 2019ApJ...877..151W}. Hence, apart from applying this test for the oscillatory nature in sector 72, we made segment-wise LSP analyses of the source's TESS LCs. 
 We fit the PSD of the respective LC using minimization of negative likelihood method to derive the free parameters of the fitting PSD model. The joint likelihood function ($\mathcal{L}$) is given as
 \begin{equation}
    \label{lglkd}
    \mathcal{L} = -2\sum_{j}\frac{I_j}{P_j}+\text{log }P_j
\end{equation}
where $I_j$ and $P_j$ denote the observed periodogram and the model spectrum, respectively. The corresponding uncertainties of the fitting parameters were evaluated following \citet{2012A&A...544A..80G}.
\subsection{Phase Dispersion Minimization (PDM)}
Apart from the PSD investigation, which is a Fourier-based approach, PDM is another tool well-suited for searching for any periodic variations in a LC \citep{1978ApJ...224..953S}. It has the ability to handle an unevenly sampled dataset with varying time gaps. Additionally, it is effective with sparse data and is particularly useful for non-sinusoidal variations. It minimizes the dispersion of a dataset at constant phase, defined by the statistic $\Theta=s^2/\sigma^2$, where $\sigma^2$ is the variance of the dataset. Here we take $M$ distinct samples of the dataset, having similar phases with variances $s_j^2\ (j=1, ..., M)$ and $n_j$ data points, and $s^2$ is the overall  variance of the samples, given as 
\begin{equation}
    s^2=\frac{\sum(n_j-1)s_j^2}{\sum n_j-M}.\\
\end{equation}
The goal of this method is to minimize the variance of the dataset over different trial periods with respect to the mean LC. We have utilized {\tt PyAstronomy} \citep{2019ascl.soft06010C} python package to estimate the $\Theta$-statistic at different trial frequencies. 
 \section{Results}
\begin{figure*}
    \centering
    \includegraphics[width=0.47\linewidth, trim=0 0 0.4cm 0, clip]{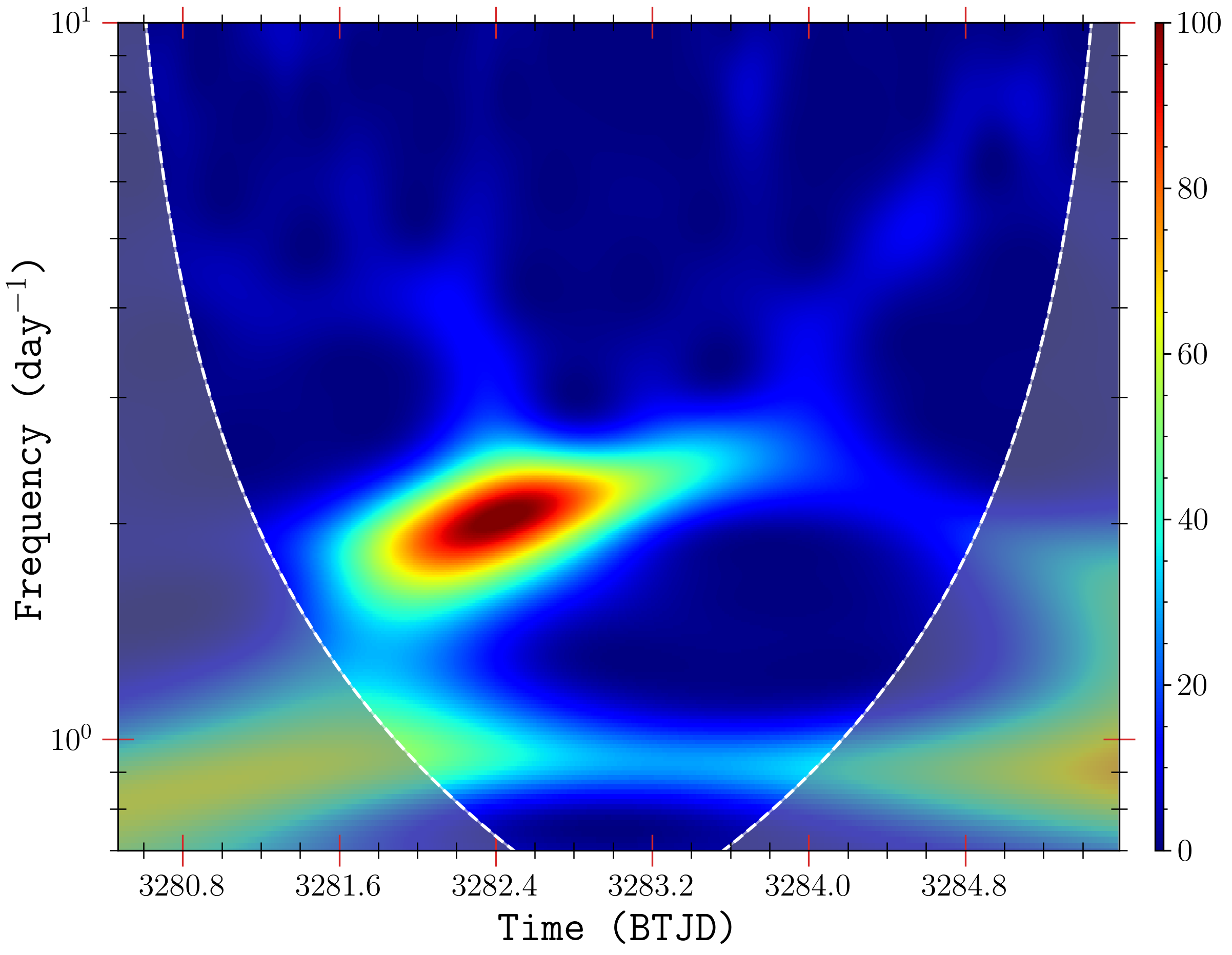}
    \includegraphics[width=0.47\linewidth, trim=0 0 0 -1cm, clip]{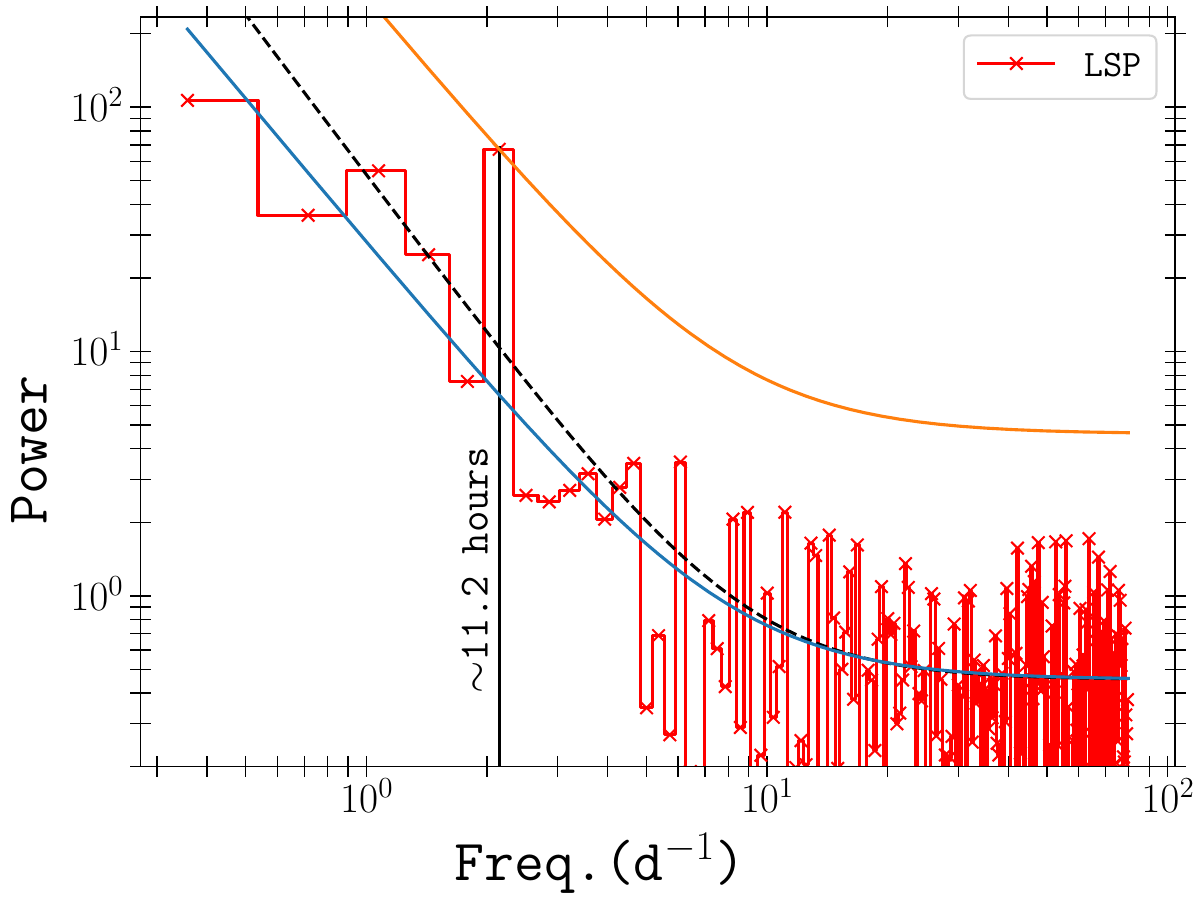}
    \includegraphics[width=0.47\linewidth, trim=0.5cm 0 0.cm 0cm, clip]{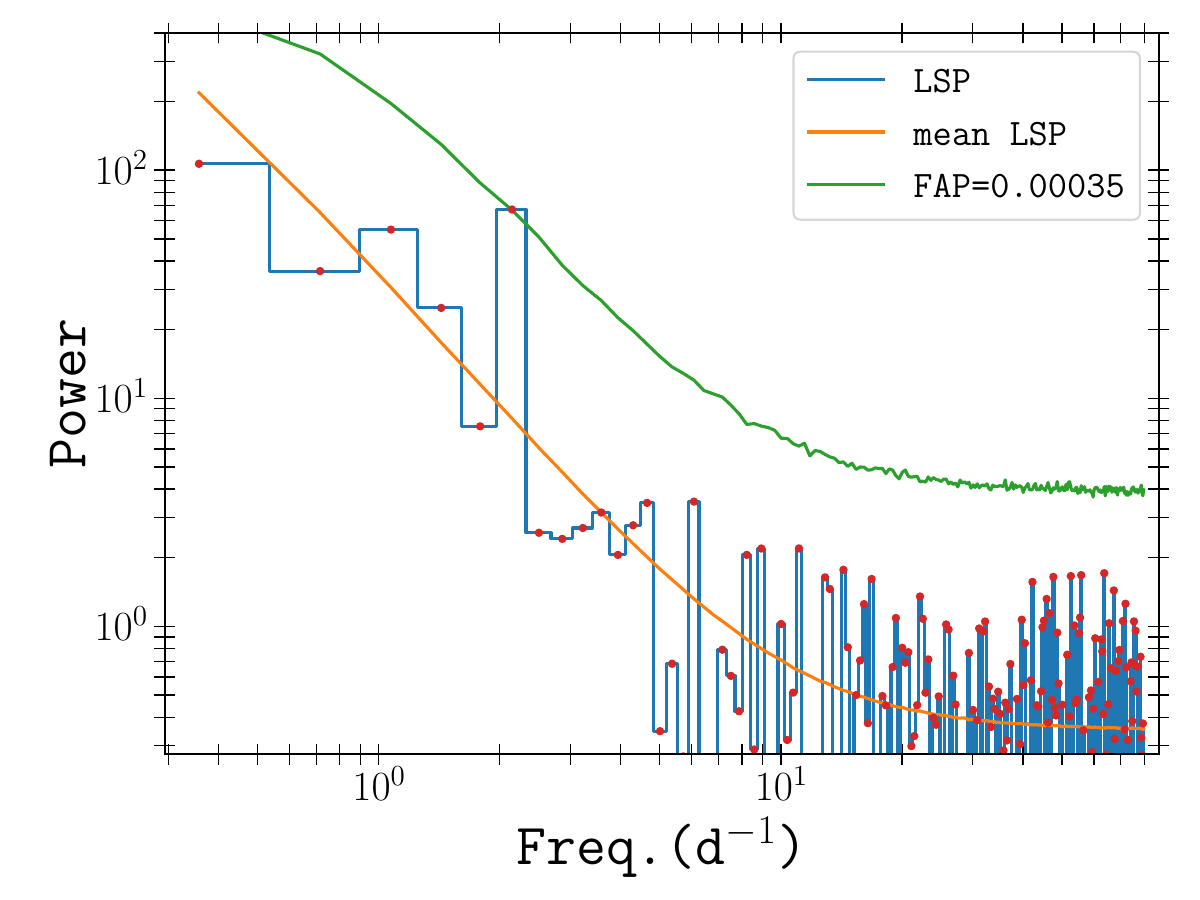}
    \includegraphics[width=0.47\linewidth, trim=0.cm 0 0.3cm 0cm, clip]{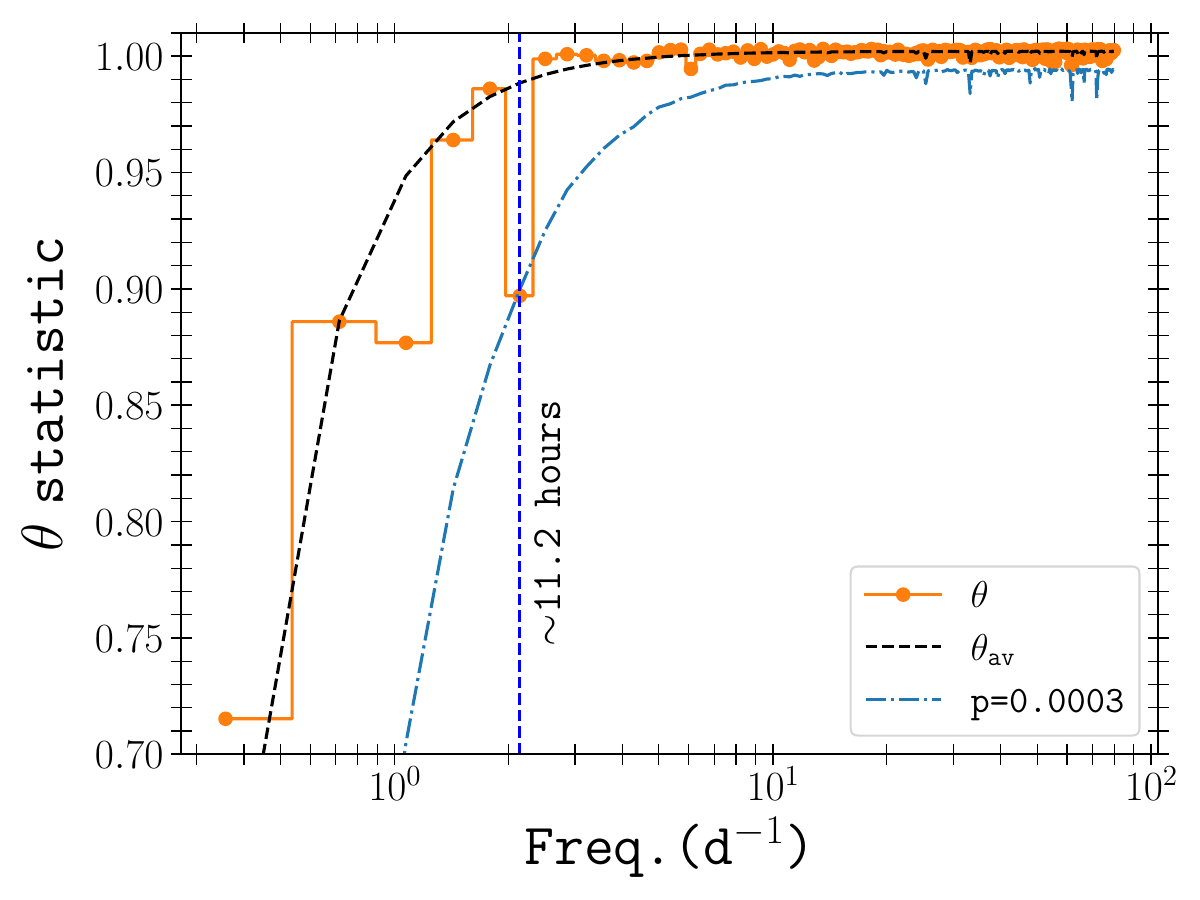}
    \caption{Upper left panel: WWZ plot for the epoch 3280.47 -- 3285.65 BTJD, showing the apparent presence of a transient QPO signature; the mildly shaded portion below the white dashed lines highlights the ROI. Upper right panel: LSP plot for the epoch 3281.3 -- 3284.1 BTJD, where the PSD shows a strong peak at $\sim11.2$ hours. The black dashed, and the blue lines respectively denote the LSP fitting with and without taking the peak frequency of $\sim2.2~\text{d}^{-1}$ into account. The QPO peak is found to be at around $4.11\sigma~\text{local significance and}~3.06\sigma$  global significance levels as indicated by the orange line. Lower left panel: Similar to the upper right panel, where we first simulated $10^5$ LCs having similar statistical features as those of LC between the epochs mentioned above, followed by computation of LSPs of those LCs; the LSP peak is found at false alarm probability (FAP) of 0.00035 ($\sim$3.58$\sigma$). Lower right panel: Phase dispersion minimization analysis showing a significant dip in that statistic. The mean statistic and the percentile confidence $p$-value have been estimated from simulating $10^5$ LCs and evaluating their individual PDMs, similarly to the treatment of the LSP.}
    \label{fig:4}
\end{figure*}
\begin{figure*}
    \centering
    \includegraphics[width=0.75\linewidth]{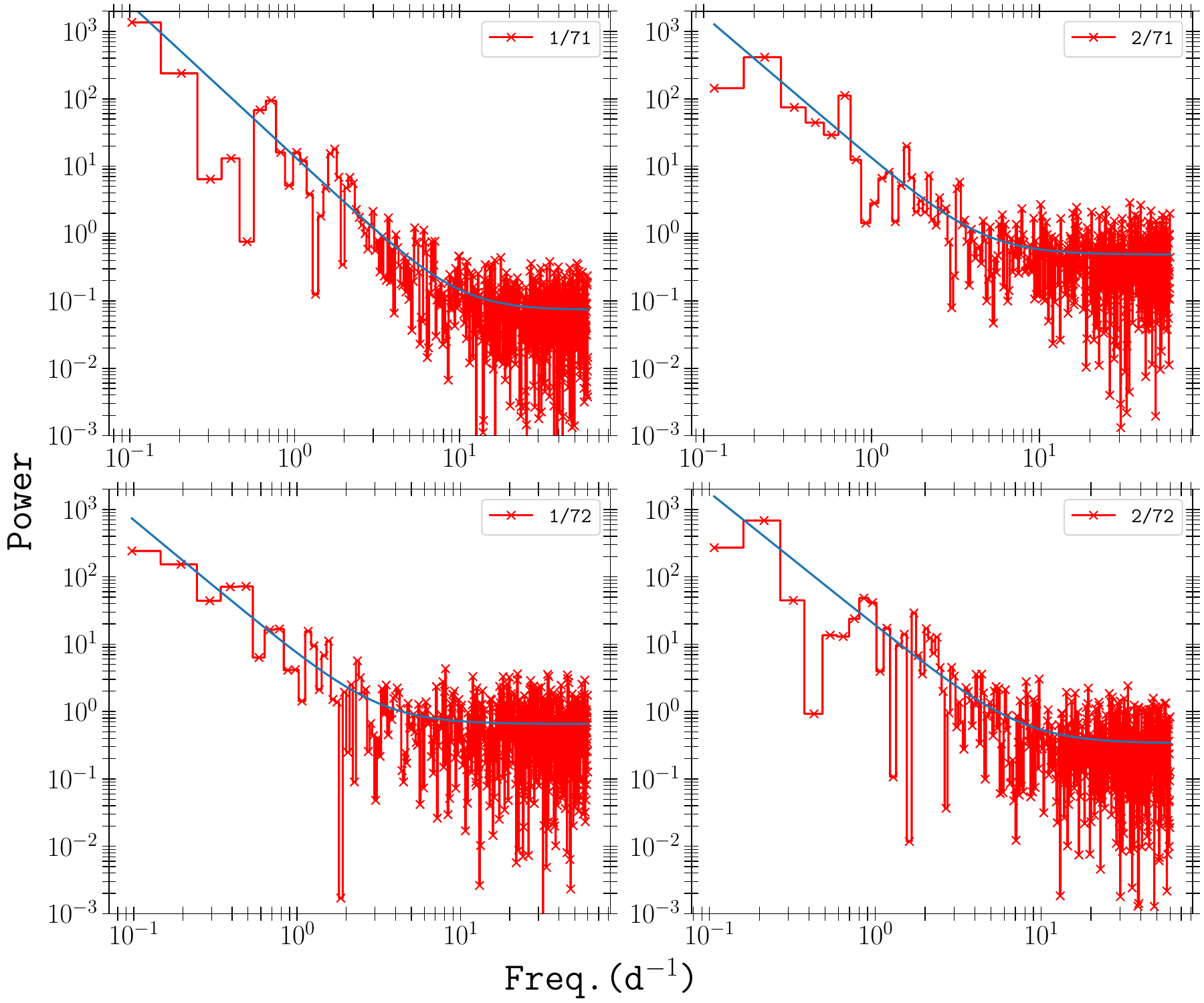}
    \caption{Segmentwise LSP plots of PKS 0735+178, along with their fitting with eq.~\ref{power_law}; the labels are in the form segment/sector.}
    \label{fig:5}
\end{figure*}
 \subsection{QPO features}
 \noindent
 Looking at the WWZ map in Fig.~\ref{fig:4}, a region of high power consistent with a transient QPO signal can be observed at  $\approx2.04~d^{-1}$, with a maximum around BTJD~3282.43. We then computed the LSP of the portion of the LC in sector 72 between BTJD~3281.3 and 3284.1 to try to confirm the existence of the QPO feature and to determine the local and global significance levels of the prominent frequency peak. The lower panel of Fig.~\ref{fig:4} includes the periodogram of this selected portion of the LC, which can be seen to be well described by a simple power-law at lower frequencies with Poisson noise dominating at higher frequencies. The peak with the greatest power is at $2.15~d^{-1}$. To estimate its significance, we implemented two distinct approaches to find the intrinsic PSD shape during fitting the periodogram, first including the peak frequency and then excluding it, employing the negative likelihood minimization method to derive the free parameters following \cite{2010MNRAS.402..307V}. The two sets of parameters were further tested for their goodness of fit using the Bayesian Information Criterion (BIC) approach, which relies on the negative likelihood minimization method and is given as 
 \begin{equation}
    BIC=k\cdot\text{ln}(n)~-~\mathcal{L}
\end{equation}
 where $k~\text{and}~n$ respectively are the number of free parameters and the number of frequency points. We find that excluding the peak frequency bin during the fit offers a lower $BIC$ value for PSD shape, hence it is considered here a better fit than the one that includes it. The determination of significance levels corresponding to a model spectrum requires consideration of the number of independent frequencies that are examined and is given by \citep[see Eq.~(16) and related discussion in][]{2005A&A...431..391V}  
 \begin{equation}
     \gamma_\epsilon=-2~ln~[1-(1-\epsilon_{n'})^{1/n'}],
 \end{equation}
 where $\epsilon_{n'}$ is the probability values (for 95$\%$ significance level, $\epsilon_{n'}=0.05$), and  $n'$ is the number of independent trials or frequencies. The PSD shape in the lower panel of Fig.~\ref{fig:4} is found to be dominated by Poisson noise beyond $\sim10~d^{-1}$, so we conservatively put a nominal cutoff frequency of $20~d^{-1}$ to set $n'~(\text{set to }55)$. Though higher values of $n'$ reduce the global significance of any prominent peak, with this choice, following \cite{2005A&A...431..391V}, we found that the LSP gives a reliable QPO feature of period $\sim 11.2$~h at around $4.11\sigma~\text{and}~3.06\sigma$ local and global significance levels, respectively.\\\\
 We also considered an alternative method to find the significance of the peak, by performing simulations of many similar LCs following \cite{2013MNRAS.433..907E}. We simulated a total of $10^5$ LCs having similar PSD shapes and probability distribution functions (PDFs) as the input LC.  We then estimated the LSPs of those LCs to obtain a distribution of powers at each temporal frequency. The 11.2~hr peak of the LSP was found to have a false alarm probability (FAP) of around 0.00035, as shown in the lower left panel of Fig.~\ref{fig:4}.
 The PDM analysis of the same dataset is presented in the lower right panel of Fig.~\ref{fig:4}, where we find a significant dip in the $\Theta$ statistic near an 11.2~hr period. To estimate the false alarm probability levels, we calculated the $\Theta$-statistic of all the $10^5$ simulated LCs, similarly to the way we did for the LSPs. The 11.2~hr signal has a FAP ($p$-value) $\approx 0.0003$.
 \subsection{Segmentwise light curve features}
 \noindent
 Over the two consecutive TESS sectors, PKS 0735+178 exhibited both variable and quiescent phases.
 Segment 1/71 has an overall variability amplitude of $\sim3\%$, though the flare region (C1 and C2) shows an amplitude of $>5\%$, pointing to a relaxation in variability after the modest double flare. The quiescent phase continues in segment 2/71, after which the source exhibits greater variability throughout sector 72, reaching an amplitude of $\sim6.5\%$ in segment 2/72, where, near the end of the observations, it shows a transient QPO signature of period $\sim11.2$ hours. Fig.~\ref{fig:5} presents the PSDs of the four individual segments, along with their red noise fits. Table~\ref{tab:2} includes the parameters of these segmentwise PSDs sectors 71 and 72. Over the four segments, the source depicts a PSD slope of $\gtrsim2$, which is typical of optical emission from blazars \citep[e.g.][]{2019ApJ...877..151W}.
 \subsection{Comparison of $\gamma-$ray data with optical data}
 \noindent
 The $\gamma-$ray LC in Fig.~\ref{fig:3} shows that the source was in a rather quiescent phase throughout the epochs of Sectors 71 and 72. During the flare and shortly before (but not during) the QPO timespan, modest increases in the one-day binned $\gamma-$ray flux appear to be present; however, the uncertainties are quite large.  Hence, we cannot compute excess variances or variability timescales, nor conduct periodicity tests, on these Fermi-LAT data. 
 \section{Discussion and Conclusion}
 \noindent
 In this paper, we assessed the optical behavior of the blazar PKS~0735+178, observed in sectors 71 and 72 with TESS. The source showed a modest flare with apparently two components (C1 and C2) in segment 1/71, followed by a sharp decrease in its variability thereafter in Sector 71. While C1 appears quite symmetric, C2 is clearly asymmetric, with the decay slower than the rise. The source became more variable toward the end of segment 1/72, and near the end of segment 2/72, it exhibited a strong QPO feature of a period of around 11.2 hours. We employed WWZ, LSP, and PDM to analyze the LCs. The flare spans a total of around 4.6 days, with individual components of 1.25 days and 3.35 days.  The QPO has local and global significance levels of around $4.11\sigma \text{ and } 3.06\sigma$ with methods described in \cite{2005A&A...431..391V}. Following \cite{2013MNRAS.433..907E}, the peak was found to be at $\sim$3.58$\sigma$. Through a non-sinusoidal analysis approach with the PDM, the QPO feature appears at $\sim$3.62$\sigma$. \\
 \\ 
There are two major classes of emission models for AGN variability: those involving shocks or instabilities in jets and those invoking instabilities on or above accretion disks. The latter class must dominate when jets are weak or nonexistent, while the former is expected to predominate in blazars \citep[e.g.][and references therein]{1993ApJ...406..420M, 1995ARA&A..33..163W, 1995PASP..107..803U, 2019ARA&A..57..467B, 2025A&ARv..33....8R}. QPOs can be associated with the spin of the central SMBH of the blazar, resulting in instabilities that cause the accretion disk to precess and/or produce bright hot spots. The presence of a single dominant hot spot on the accretion disk close to the innermost stable circular orbit allowed by general relativity or pulsational modes in the disk, are the most simple explanations for the observation of QPOs in blazars on IDV timescales \citep[e.g.][]{1991A&A...245..454A, 1993ApJ...406..420M, 1993ApJ...411..602C, 2008ApJ...679..182E}. \\
\\
The turbulence behind a shock traveling down a jet can create dominant eddies, whose turnover times can result in QPOs in emission at various EM bands \citep[e.g.][]{1992vob..conf...85M, 2014ApJ...780...87M, 2016ApJ...820...12P}. Additionally, QPOs may result from a relativistic shock propagating along a helical jet, which can be caused by precession or magnetohydrodynamical instabilities \citep{1999ApJ...524..650H}. When a relativistic shock travels down such a perturbed jet, it will cause noticeably increased emission at the points where it meets an area of enhanced electron density and/or magnetic field that corresponds to such a non-axisymmetric structure \citep[e.g.][]{1992A&A...255...59C, 1992A&A...259..109G}. The observer can notice significant variations in the observed jet emission because Doppler boosting is a sensitive function of viewing angle.\\
\\
The low variability in the $\gamma-$ray LC, with only modest rises in flux levels during the flare and near the QPO, may help constrain the location of the emission region of the observed flares and QPO features.  This apparently weak correlation may indicate that the origins of the optical and $\gamma-$ray emissions are not necessarily at the same location. \\
\\
 Estimation of the central black hole mass of BLLs, including PKS~0735+178, is quite difficult because of the absence of any broad spectral lines and because blazars are nearly face-on oriented. Thus, traditional primary methods of black hole mass estimation, such as stellar gas kinematics, reverberation mapping, and megamaser kinematics, cannot be employed as they are applicable only when one has, respectively, high-resolution spectroscopy of the host galaxy, detection of high ionization emission lines from the broad lines, or the source is oriented in an edge-on geometry. Other than these primary methods, secondary methods to deduce the masses include the assessment of the bulge velocity dispersion or luminosity, or the S\'{e}rsic index \citep[e.g.][]{2004ASPC..311...69V, 2013ApJ...773...90G, 2021MNRAS.504.5188H, 2026ApJ..1000...48B}, but they are also not applicable to PKS~0735+178.\\
\\
\cite{2009ApJ...690..216G} offered a means to estimate the black hole mass of an AGN using the period of a QPO signature, assuming that it originates from the orbital motion of some blob near the inner regions of the accretion disk \citep[e.g.][]{1993ApJ...406..420M}, which is nominally taken as the marginally stable orbit, although emission from infalling matter at even smaller radii may be possible \citep{1982Natur.300..506A}.
In this case, one can relate mass to periodicity via
 \begin{equation}
     M=\frac{3.23\times10^4~P}{(r^{3/2}+a)(1+z)}M_\odot~,
 \end{equation}
 where $P,~r,~a,\text{ and }z$ are period (in seconds), radial distance (here set to the marginally stable orbit in units of gravitational radius: $GM/c^2$), spin parameter (often taken as 0 and 0.9982 for Schwarzschild and maximal Kerr black holes, respectively), and redshift of the source, respectively. Then the estimated mass of PKS~0735+178 is $(6.10\pm1.05)\times10^7~M_\odot\text{ and }(3.88\pm0.67)\times10^8~M_\odot$ for a spinless and a maximally rotating central supermassive black hole, respectively. An earlier estimation of the central SMBH mass of this blazar is $\approx 1.89 \times \ \rm{10}^{8} \ \rm{M}_{\odot}$ (corresponding to the Eddington luminosity L$_{Edd} = \rm{2.5} \times \ \rm{10}^{46} \ \rm{erg} \ {s}^{-1}$), using the optical intraday variability timescale and adopting a jet Doppler factor $\delta \approx$ 3.5 \citep{2012NewA...17....8G}. This mass estimation is in the range of the present study.  In a more realistic situation for a blazar, the fluctuations produced in the disk would be advected into the jet and amplified by Doppler boosting.  In that case, the observed period would be reduced by a factor of $\delta$ from the inherent one, indicating that the actual black hole mass is actually increased by that $\delta$ factor.  Then these two mass estimates are only compatible if the spin is small or the region of enhanced emission is further out than the marginally stable orbit.
\section*{ACKNOWLEDGMENTS}
\noindent
 This paper includes data collected with the TESS mission, obtained from the MAST data archive at the Space Telescope Science Institute (STScI). Funding for the TESS mission is provided by the NASA Explorer Program. STScI is operated by the Association of Universities for Research in Astronomy, Inc., under NASA contract NAS 5–26555. This work also made use of data from the Fermi Science Support Center (FSSC). The \textit{Fermi}-LAT gamma-ray data analysis was performed on the TDLI-Astro cluster at Tsung-Dao Lee Institute.\\
\\
{\it Data Availability:} The optical dataset from the TESS, including the cutouts of the FFI of sectors 71 and 72, can be found in MAST through \url{https://mast.stsci.edu/tesscut/}. Additionally, the derived optical and $\gamma$-ray light curve data can be made available upon reasonable request.\\
\\
{\it Software:} lightkurve \citep{2018ascl.soft12013L}, 
SciPy \citep{2020SciPy-NMeth}, PyAstronomy \citep{2019ascl.soft06010C}, {Fermipy (V1.0.1; \url{https://fermipy.readthedocs.io/en/latest/} ;\citet{Fermipy_Version}).
\bibliography{ref} 
\bibliographystyle{mnras}
\section*{appendix}
    \begin{figure}
    \centering
    \includegraphics[width=0.95\linewidth]{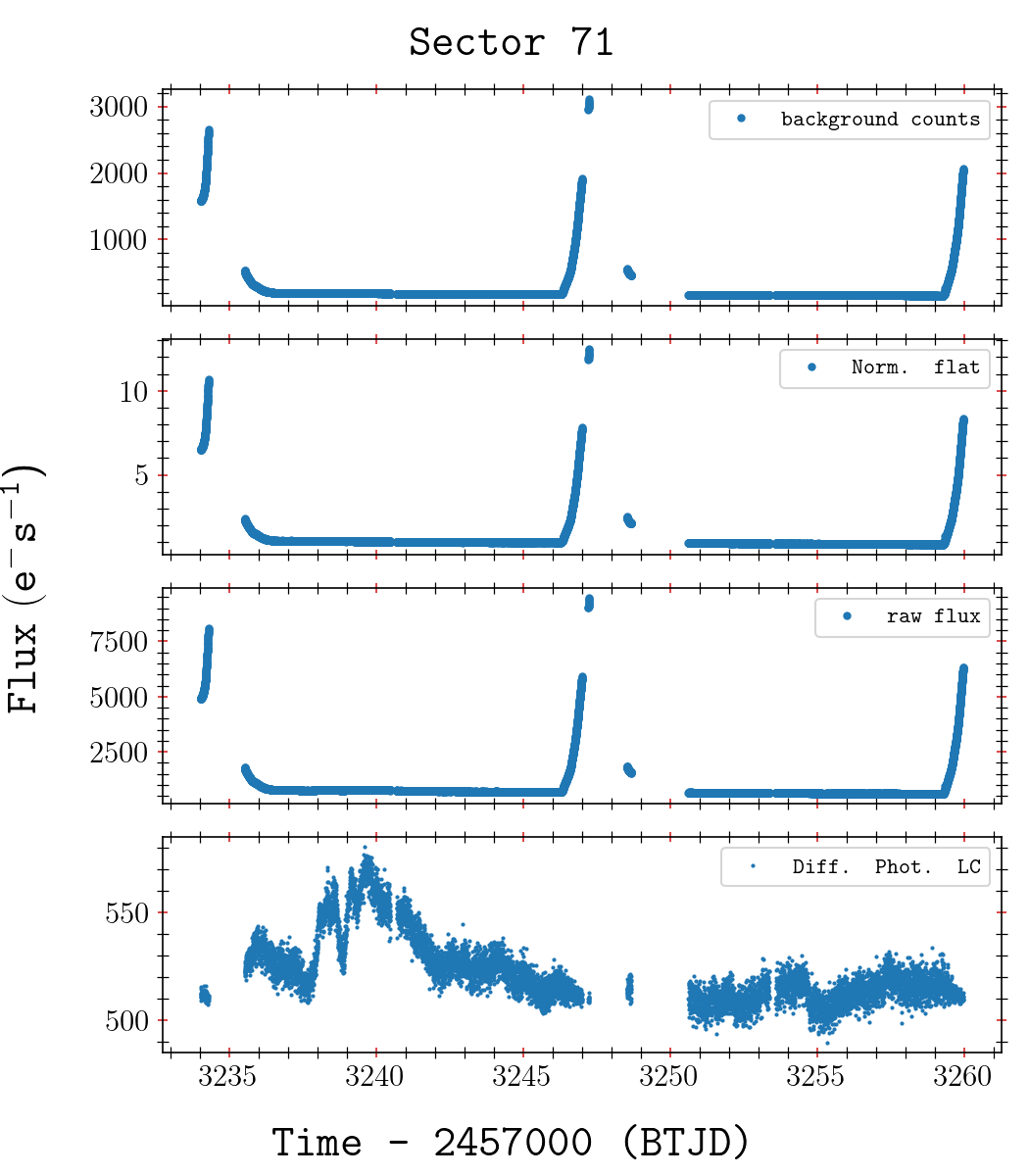}
    \caption{First panel shows the background LC, second panel shows the normalized LC of `D' or the normalized flat or the pixel response function, third panel includes the raw LC of the source (`S'), fourth panel depicts the unnormalized pixel response function corrected LC of the source; all these plots correspond to sector 71; the pixel response function show significant variations from unity, so epochs for which they are between 0.85 to 1.15 (15\% cuts) only were considered for correction of source raw flux; additionally epochs where the normalized flat folllows a sharp increasing trend (near the end of each segments) were also discarded during LC analyses.}
    \label{fig:Ap1}
    \end{figure}
    \begin{figure}
    \centering
    \includegraphics[width=0.95\linewidth]{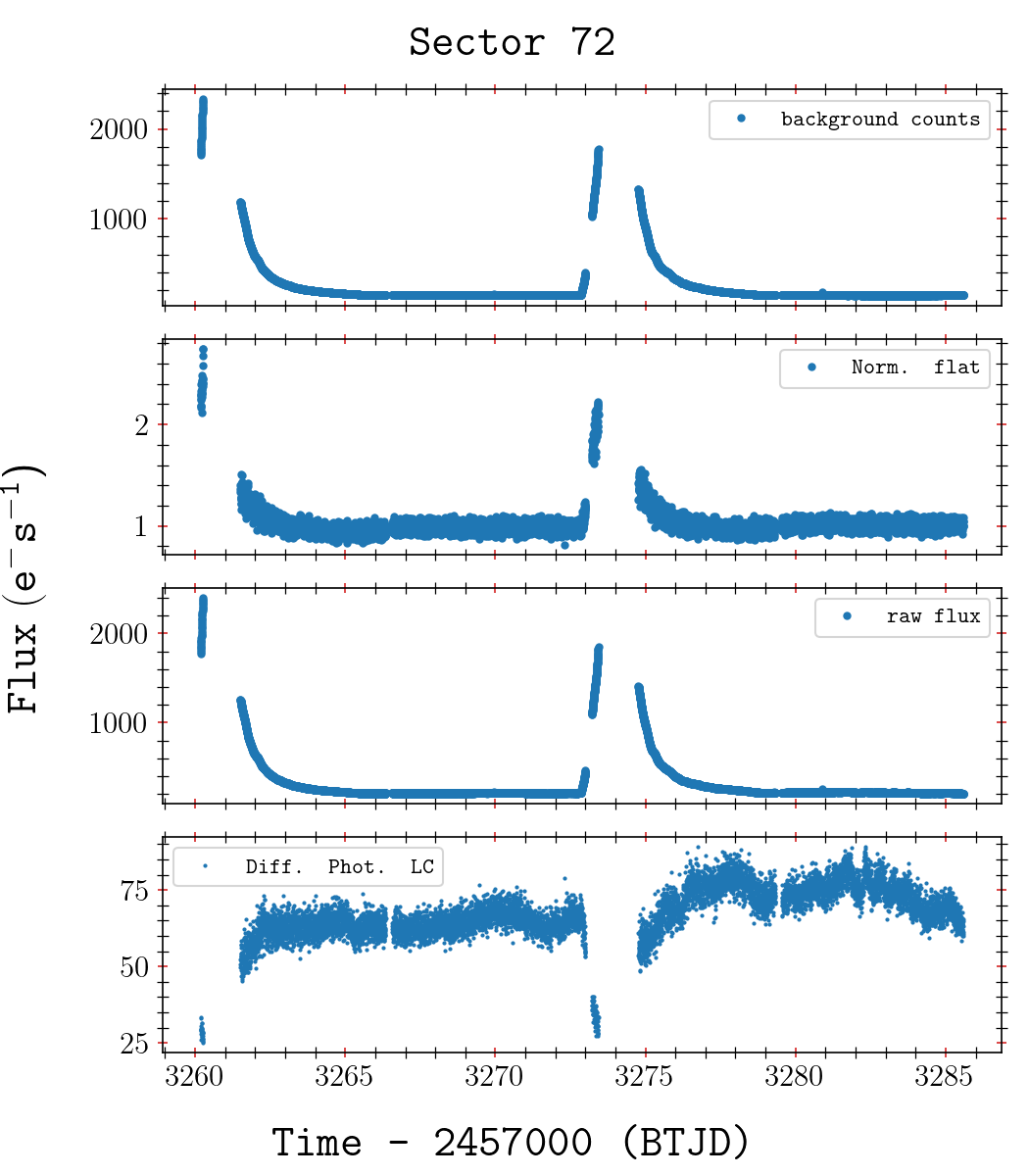}
    \caption{Same as Fig.~\ref{fig:Ap1}, but for sector 72.}
    \label{fig:Ap2}
    \end{figure}
\bsp
\label{lastpage}
\end{document}